\documentclass[prl,showpacs,showkeys,byrevtex,twocolumn]{revtex4}%
\usepackage{amsfonts}
\usepackage{amsmath}
\usepackage{amssymb}
\usepackage{graphicx}%
\providecommand{\U}[1]{\protect\rule{.1in}{.1in}}

\begin{document}
\preprint{ }
\title[ ]{Coherent Communication with Continuous Quantum Variables}
\author{Mark M. Wilde$^{1}$}
\email{mark.wilde@usc.edu}
\author{Hari Krovi$^{2}$}
\author{Todd A. Brun$^{2}$}
\affiliation{$^{1}$Signal and Image Processing Institute, $^{2}$Communication Sciences
Institute, Department of Electrical Engineering, University of Southern
California, Los Angeles, California 90089, USA}
\keywords{coherent communication, cobit channel, conat channel, continuous variables}
\pacs{03.67.-a, 03.67.Hk, 42.50.Dv}

\begin{abstract}
The coherent bit (cobit) channel is a resource intermediate between classical
and quantum communication. It produces coherent versions of teleportation and
superdense coding. We extend the cobit channel to continuous variables by
providing a definition of the coherent nat (conat) channel. We construct
several coherent protocols that use both a position-quadrature and a
momentum-quadrature conat channel with finite squeezing. Finally, we show that
the quality of squeezing diminishes through successive compositions of
coherent teleportation and superdense coding.

\end{abstract}
\volumeyear{2007}
\volumenumber{ }
\issuenumber{ }
\eid{ }
\date{\today}
\received{\today}

\revised{}

\accepted{}

\published{}

\startpage{1}
\endpage{ }
\maketitle


The coherent bit (cobit) channel is a useful resource for quantum
communication with discrete variables (DV)\ \cite{prl2004harrow}. The cobit
channel $\Delta_{\sigma_{Z}}$ copies $\sigma_{Z}$ eigenstates coherently from
Alice to Bob: $\left\vert i\right\rangle ^{A}\rightarrow\left\vert
i\right\rangle ^{A}\left\vert i\right\rangle ^{B}$. \textquotedblleft
Coherence\textquotedblright\ in this context is synonymous with
linearity---the maintenance and linear transformation of superposed states. We
name the cobit channel $\Delta_{\sigma_{Z}}$ the \textit{Pauli-}$Z$\textit{
cobit channel}. One can similarly define the \textit{Pauli-}$X$\textit{ cobit
channel} $\Delta_{\sigma_{X}}$ that coherently copies $\sigma_{X}$ eigenstates.

In this paper, we extend the notion of the cobit channel to
continuous-variable (CV)\ quantum information processing
\cite{book2003braunstein,revmod2005braunstein}. We name the CV version the
\textit{conat} channel in analogy with Shannon's name for the information in a
continuous random variable measured in units of the natural logarithm. We then
construct several coherent protocol primitives. We lastly address
\textit{duality under resource reversal} and discover a difference between the
DV\ and CV\ coherent channels due to finite squeezing.

What is coherent communication useful for? Insights into quantum protocols
occur by replacing classical bits with cobits---replacing a measurement and
feedforward classical communication with a coherent channel and replacing a
conditional unitary with a controlled unitary. Coherent teleportation and
superdense coding for DVs are dual under resource reversal
\cite{prl2004harrow,devetak:140503}. Two protocols are \textit{dual under
resource reversal} if one protocol generates the same resources that the other
protocol consumes and vice versa. \textit{Coherent} remote state preparation
(RSP)\ requires less entanglement than standard RSP \cite{prl2004harrow}.
Replacing classical bits with cobits produces coherent versions of several
quantum information theory protocols \cite{prl2004dev,arx2005dev}. Coherent
communication also provides an alternate construction of the newly discovered
entanglement-assisted quantum error correcting codes
\cite{science2006brun,arx2006brun}.

We structure this Letter by first motivating and providing a general
\textit{Heisenberg-representation} definition of a position-quadrature (PQ)
conat channel\ and momentum-quadrature (MQ) conat channel. We then construct
examples of CV\ coherent protocols. Finally, we analyze the duality of
coherent teleportation and coherent superdense coding under resource reversal.
We find that finitely-squeezed coherent CV teleportation and superdense coding
are dual under resource reversal only for some maximum number of compositions;
beyond that point, classical operations suffice to implement the effective
protocol. Duality does not hold when the number of compositions exceeds the maximum.

The cobit-channel definition\ immediately tempts one to define an ideal PQ
conat channel as the quantum-feedback operation $\Delta_{X}$\ which copies
position eigenstates: $\left\vert x\right\rangle ^{A}\rightarrow\left\vert
x\right\rangle ^{A}\left\vert x\right\rangle ^{B}$. The ideal MQ conat channel
is the operation $\Delta_{P}$\ that copies momentum eigenstates. We call these
conat channels \textit{ideal} because copying the eigenstates exactly requires
infinite energy.

We provide Heisenberg-representation definitions of both a finitely-squeezed
PQ conat channel $\tilde{\Delta}_{X}$ and MQ conat channel $\tilde{\Delta}%
_{P}$ as an approximation to the above ideal scenarios. The first requirement
for $\tilde{\Delta}_{X}$\ is that it approximate the ideal position-copying
behavior mentioned above. $\tilde{\Delta}_{X}$ should copy the PQ as exactly
as possible given finite squeezing. Observe the effect of the ideal PQ conat
channel $\Delta_{X}$ on a momentum eigenstate $\left\vert p\right\rangle $.
The resulting two-mode state is a maximally-entangled Bell state $\int
e^{ipx}\left\vert x\right\rangle ^{A}\left\vert x\right\rangle ^{B}\ dx$. It
is an eigenstate of the total momentum operator $\hat{p}_{A}+\hat{p}_{B}$ with
eigenvalue $p$. The second requirement is that the total momentum $\hat{p}%
_{A}+\hat{p}_{B}$ should be close to the original momentum $\hat{p}_{A}$.

An $\epsilon$\textit{-approximate PQ conat channel} $\tilde{\Delta}_{X}$
performs the following transformation with conditions:%
\begin{align}%
\begin{bmatrix}
\hat{x}_{A} & \hat{p}_{A}%
\end{bmatrix}
^{T}\  &  \underrightarrow{\tilde{\Delta}_{X}}\
\begin{bmatrix}
\hat{x}_{A^{\prime}} & \hat{p}_{A^{\prime}} & \hat{x}_{B^{\prime}} & \hat
{p}_{B^{\prime}}%
\end{bmatrix}
^{T}\\
\left[  \hat{x}_{A^{\prime}},\hat{p}_{A^{\prime}}\right]   &  =\left[  \hat
{x}_{B^{\prime}},\hat{p}_{B^{\prime}}\right]  =i\nonumber\\
\hat{x}_{A^{\prime}}  &  =\hat{x}_{A}\\
\hat{x}_{B^{\prime}}  &  =\hat{x}_{A}+\hat{x}_{\Delta_{X}}\\
\hat{p}_{A^{\prime}}  &  =\hat{p}_{A}+\hat{p}_{\Delta_{X}}\\
\left\langle \hat{x}_{\Delta_{X}}\right\rangle  &  =\left\langle \hat
{p}_{\Delta_{X}}+\hat{p}_{B^{\prime}}\right\rangle =0\\
\langle\hat{x}_{\Delta_{X}}^{2}\rangle,  &  \ \ \langle\left(  \hat{p}%
_{\Delta_{X}}+\hat{p}_{B^{\prime}}\right)  ^{2}\rangle\leq\epsilon
\end{align}
The momentum quadrature $\hat{p}_{B^{\prime}}$ is arbitrary as long as it
obeys the above constraints. An $\epsilon$\textit{-approximate MQ conat
channel} $\tilde{\Delta}_{P}$ performs the following transformation with
conditions:%
\begin{align}%
\begin{bmatrix}
\hat{x}_{A} & \hat{p}_{A}%
\end{bmatrix}
^{T}\  &  \underrightarrow{\tilde{\Delta}_{P}}\
\begin{bmatrix}
\hat{x}_{A^{\prime\prime}} & \hat{p}_{A^{\prime\prime}} & \hat{x}%
_{B^{\prime\prime}} & \hat{p}_{B^{\prime\prime}}%
\end{bmatrix}
^{T}\label{eqn:cond-m-conat-1}\\
\left[  \hat{x}_{A^{\prime\prime}},\hat{p}_{A^{\prime\prime}}\right]   &
=\left[  \hat{x}_{B^{\prime\prime}},\hat{p}_{B^{\prime\prime}}\right]
=i\nonumber\\
\hat{p}_{A^{\prime\prime}}  &  =\hat{p}_{A}\nonumber\\
\hat{p}_{B^{\prime\prime}}  &  =\hat{p}_{A}+\hat{p}_{\Delta_{P}}\nonumber\\
\hat{x}_{A^{\prime\prime}}  &  =\hat{x}_{A}+\hat{x}_{\Delta_{P}}\nonumber\\
\left\langle \hat{p}_{\Delta_{P}}\right\rangle  &  =\left\langle \hat
{x}_{\Delta_{P}}+\hat{x}_{B^{\prime\prime}}\right\rangle =0\nonumber\\
\langle\hat{p}_{\Delta_{P}}^{2}\rangle,  &  \ \ \langle\left(  \hat{x}%
_{\Delta_{P}}+\hat{x}_{B^{\prime\prime}}\right)  ^{2}\rangle\leq
\epsilon\nonumber
\end{align}
The position quadrature $\hat{x}_{B^{\prime\prime}}$ is arbitrary as long as
it obeys the above constraints. We require $0<\epsilon<1$ for both conat channels.

Fourier transformation gives the relationship between a MQ\ and PQ conat
channel:\ $\tilde{\Delta}_{P}=\left(  \mathbb{F}^{-1}\otimes\mathbb{F}%
^{-1}\right)  \ \tilde{\Delta}_{X}\ \mathbb{F}$. Both a PQ\ and MQ\ conat
channel implement coherent teleportation---just as Braunstein and Kimble use
both PQ\ and MQ homodyne detection in their teleportation scheme
\cite{prl1998braunstein}. Our coherent teleportation protocol for CVs is
similar to theirs except that a PQ and MQ conat channel replaces the
feedforward classical communication and the PQ and MQ homodyne measurement respectively.

Coherent teleportation protocols using the above $\epsilon$-approximate conat
channels have a coherent-state-averaged fidelity greater than the
preparation-and-measurement limit \cite{prl1998braunstein}\ of $1/2$ if
$\epsilon<1$. We consider this limit of $1/2$ as opposed to other limits
\cite{PhysRevA.64.010301,caves:040506}\ because surpassing this limit implies
the use of an entangled resource for teleporting an arbitrary coherent state.
We simply wish to relate the measure of conat-channel performance to the
presence of entanglement.

We can measure PQ conat-channel performance by sending its two output modes
through a 50:50 beamsplitter and determining the second moments of one
output's PQ and the other output's MQ. Both outputs---the relative position
and total momentum---should have second moment bounded by $\epsilon$ and
$\left\langle \hat{p}_{A}^{2}\right\rangle +\epsilon$ respectively in order to
be an $\epsilon$-approximate PQ conat channel.

The above definitions are sufficient for implementing a coherent teleportation
with CVs via conat channels. They are also necessary for realizing two conat
channels as a result of a coherent superdense coding.

We define a two-mode system with Heisenberg-picture quadrature operators
$\hat{x}_{A}$, $\hat{p}_{A}$, $\hat{x}_{B}$, $\hat{p}_{B}$\ as $\epsilon
$\textit{-position-correlated} if $\langle\left(  \hat{x}_{A}-\hat{x}%
_{B}\right)  ^{2}\rangle\leq\epsilon$ and $\langle\left(  \hat{p}_{A}+\hat
{p}_{B}\right)  ^{2}\rangle\leq\epsilon$. It is\textit{ }$\epsilon
$\textit{-momentum-correlated} if $\langle\left(  \hat{x}_{A}+\hat{x}%
_{B}\right)  ^{2}\rangle\leq\epsilon$ and $\langle\left(  \hat{p}_{A}-\hat
{p}_{B}\right)  ^{2}\rangle\leq\epsilon$. It is $\epsilon$%
\textit{-position-entangled} or $\epsilon$\textit{-momentum-entangled}
if$\ \epsilon<1$ \cite{prl2000lmduan}.%

\begin{figure}
[ptb]
\begin{center}
\includegraphics[
height=1.1969in,
width=3.3451in
]%
{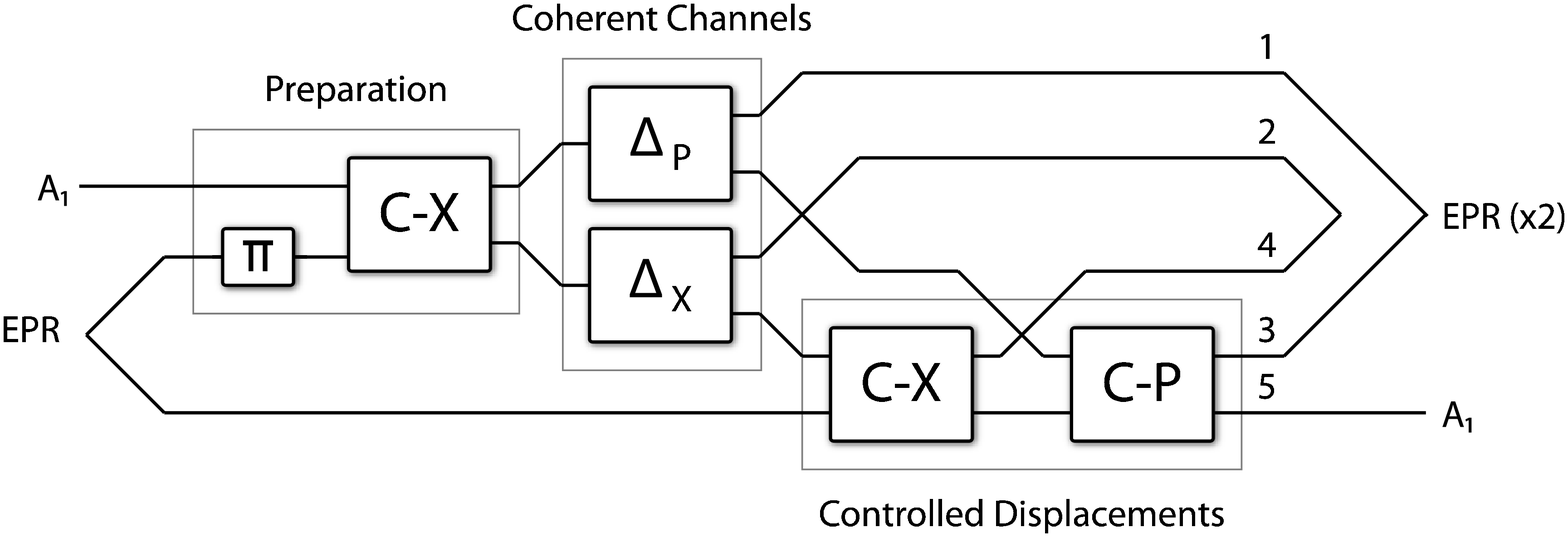}%
\caption{Coherent teleportation. $\pi$ is a reflection operation. C-X and
C-P\ are a controlled-position displacement and a controlled-momentum
displacement respectively.}%
\label{fig:coh-tele-harrow}%
\end{center}
\end{figure}
We use several operations throughout this paper. A \textit{reflection}
reverses the quadrature operators of a single mode: $\hat{x}\rightarrow
-\hat{x}$, $\hat{p}\rightarrow-\hat{p}$. A \textit{controlled-position
displacement} is a two-mode operation: $\hat{x}_{1}\rightarrow\hat{x}_{1}$,
$\hat{p}_{1}\rightarrow\hat{p}_{1}-\hat{p}_{2}$, $\hat{x}_{2}\rightarrow
\hat{x}_{2}+\hat{x}_{1}$, $\hat{p}_{2}\rightarrow\hat{p}_{2}$. A
\textit{controlled-momentum displacement} is as follows: $\hat{x}%
_{1}\rightarrow\hat{x}_{1}-\hat{x}_{2}$, $\hat{p}_{1}\rightarrow\hat{p}_{1}$,
$\hat{x}_{2}\rightarrow\hat{x}_{2}$, $\hat{p}_{2}\rightarrow\hat{p}_{2}%
+\hat{p}_{1}$.

Vaidman provided the first theoretical description of CV\ teleportation
\cite{PhysRevA.49.1473}\ followed by Braunstein and Kimble's in terms of the
quadratures of the EM\ field \cite{prl1998braunstein}. We construct a coherent
version of Braunstein and Kimble's protocol: \textit{coherent teleportation}
(Fig.~\ref{fig:coh-tele-harrow}). The protocol is similar to Harrow's for DVs
\cite{prl2004harrow}. Suppose Alice wants to teleport a quantum state $A_{1}$
to Bob coherently via PQ and MQ conat channels. They possess two modes $A$ and
$B$ that are $\epsilon_{1}$-position-correlated with the additional
restriction that they have mean-zero relative position and total
momentum:\ $\left\langle \hat{x}_{A}-\hat{x}_{B}\right\rangle =\left\langle
\hat{p}_{A}+\hat{p}_{B}\right\rangle =0$. Alice performs a reflection on her
mode $A$ followed by a controlled-position displacement on her two modes
$A_{1}$ and $A$. These two operations replace the beamsplitter in Braunstein
and Kimble's teleportation protocol. She sends her first mode\ through an
$\epsilon_{2}$-approximate MQ conat channel $\tilde{\Delta}_{P}$ and her
second mode through an $\epsilon_{3}$-approximate PQ conat channel
$\tilde{\Delta}_{X}$. The two conat channels $\tilde{\Delta}_{X}$ and
$\tilde{\Delta}_{P}$ replace the feedforward classical communication and
position-quadrature and momentum-quadrature homodyne measurements
respectively. The global state becomes a five-mode state. Alice possesses her
two original modes and Bob possesses two additional modes due to both conat
channels. Bob performs a controlled-position and controlled-momentum
displacement according to Fig.~\ref{fig:coh-tele-harrow}. These controlled
displacements replace the conditional displacements in the original protocol.
The five modes then have the Heisenberg-picture observables:%
\begin{align}
\hat{x}_{1}  &  =\hat{x}_{A_{1}}+\hat{x}_{\Delta_{P}},\ \ \hat{p}_{1}=\hat
{p}_{A}+\hat{p}_{A_{1}}\\
\hat{x}_{2}  &  =\hat{x}_{A_{1}}-\hat{x}_{A},\ \ \hat{p}_{2}=-\hat{p}_{A}%
+\hat{p}_{\Delta_{X}}\nonumber\\
\hat{x}_{3}  &  =\left(  \hat{x}_{A}-\hat{x}_{B}\right)  -\hat{x}_{A_{1}}%
+\hat{x}_{B^{\prime\prime}}-\hat{x}_{\Delta_{X}},\nonumber\\
\hat{p}_{3}  &  =\hat{p}_{A}+\hat{p}_{A_{1}}+\hat{p}_{\Delta_{P}}\nonumber\\
\hat{x}_{4}  &  =\hat{x}_{A_{1}}-\hat{x}_{A}+\hat{x}_{\Delta_{X}},\ \ \hat
{p}_{4}=\hat{p}_{B^{\prime}}-\hat{p}_{B}\nonumber\\
\hat{x}_{5}  &  =\hat{x}_{A_{1}}+\left(  \hat{x}_{B}-\hat{x}_{A}\right)
+\hat{x}_{\Delta_{X}},\nonumber\\
\hat{p}_{5}  &  =\hat{p}_{A_{1}}+\left(  \hat{p}_{A}+\hat{p}_{B}\right)
+\hat{p}_{\Delta_{P}}\nonumber
\end{align}
Bob possesses the teleported state---Alice's original mode $A_{1}$---in mode
five. The coherent-state-averaged teleportation fidelity $F$
\cite{science1998furusawa} is as follows%
\begin{equation}
F=2/\left[  \left(  \left\langle \left(  \Delta\hat{x}_{tel}\right)
^{2}\right\rangle +1\right)  \left(  \left\langle \left(  \Delta\hat{p}%
_{tel}\right)  ^{2}\right\rangle +1\right)  \right]  ^{1/2}%
\end{equation}
where $\left(  \hat{x}_{tel},\hat{p}_{tel}\right)  $ is the teleported mode. A
lower bound on the fidelity $F$ using the above coherent teleportation
protocol is $2/\left(  \left(  2+\epsilon_{1}+\epsilon_{2}\right)  \left(
2+\epsilon_{1}+\epsilon_{3}\right)  \right)  ^{1/2}$. Suppose the
$\epsilon_{1}$-position-correlated state is entangled so that $\epsilon_{1}%
<1$. Then the coherent teleportation protocol exceeds the classical limit of
$1/2$ \cite{prl1998braunstein} if both $\epsilon_{2}<1$ and $\epsilon_{3}<1$.
Alice and Bob possess an $\left(  \epsilon_{1}+\epsilon_{2}+\epsilon
_{3}\right)  $-position-correlated state shared between modes one and three.
They also possess an $\left(  \epsilon_{1}+\epsilon_{3}\right)  $%
-momentum-correlated state shared between modes two and four. Thus the
original protocol becomes coherent with the benefit of generating two sets of
entanglement correlations if $\epsilon_{1}+\epsilon_{2}+\epsilon_{3}<1$.%

\begin{figure}
[ptb]
\begin{center}
\includegraphics[
height=1.2306in,
width=3.0658in
]%
{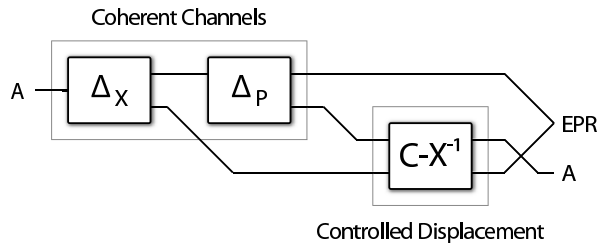}%
\caption{Alternate coherent teleportation protocol.}%
\label{fig:coh-tele-slick}%
\end{center}
\end{figure}
We provide an alternate coherent teleportation protocol which is not a direct
mapping to Kimble and Braunstein's scheme (Fig.~\ref{fig:coh-tele-slick}).
This protocol is similar to van Enk's C1 protocol \cite{pra2005vanenk}\ and to
another protocol \cite{igorprivate}. Alice possesses a mode $A$\ that she
wishes to teleport to Bob using two conjugate conat channels. Alice first
sends her mode through an $\epsilon_{1}$-approximate PQ conat channel. She
then sends her mode through an $\epsilon_{2}$-approximate MQ conat channel.
Bob possesses two modes after the two operations. He performs an inverse
controlled-position displacement on his two modes. The three modes after the
protocol have the Heisenberg-picture observables:%
\begin{align}
\hat{x}_{1} &  =\hat{x}_{A}+\hat{x}_{\Delta_{P}},\ \ \hat{p}_{1}=\hat{p}%
_{A}+\hat{p}_{\Delta_{X}}\\
\hat{x}_{2} &  =\hat{x}_{A}+\hat{x}_{\Delta_{X}},\ \ \hat{p}_{2}=\hat{p}%
_{A}+\left(  \hat{p}_{\Delta_{X}}+\hat{p}_{B^{\prime}}\right)  +\hat
{p}_{\Delta_{P}}\nonumber\\
\hat{x}_{3} &  =\hat{x}_{B^{\prime\prime}}-\left(  \hat{x}_{A}+\hat{x}%
_{\Delta_{X}}\right)  ,\ \ \hat{p}_{3}=\hat{p}_{A}+\hat{p}_{\Delta_{X}}%
+\hat{p}_{\Delta_{P}}\nonumber
\end{align}
Alice possesses the first mode and Bob possesses the last two. Mode two is the
teleported mode containing Alice's original state $A$. A lower bound on the
fidelity $F$ is $2/\left(  2+\epsilon_{1}+\epsilon_{2}\right)  $. The
teleportation fidelity exceeds the classical limit of $1/2$ if both
$\epsilon_{1}<1$ and $\epsilon_{2}<1$. Alice and Bob possess an $\left(
\epsilon_{1}+\epsilon_{2}\right)  $-momentum-correlated state shared between
modes one and three (momentum-entangled if $\epsilon_{1}+\epsilon_{2}<1$).

\begin{figure}
[ptb]
\begin{center}
\includegraphics[
height=1.209in,
width=3.339in
]%
{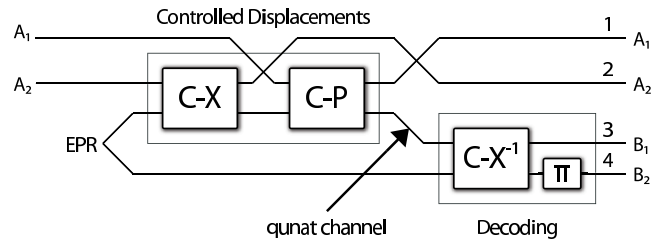}%
\caption{Coherent superdense coding implements two conat channels.}%
\label{fig:coh-superdense}%
\end{center}
\end{figure}

Braunstein and Kimble provided a theoretical proposal for a superdense coding
protocol with CVs \cite{pra2000braunstein}. They demonstrated that bipartite
CV entanglement and a qunat channel can double classical communication
capacity in the limit of large squeezing. We provide a coherent version of
their superdense coding protocol by implementing both a PQ and a MQ conat
channel rather than two classical nat channels (Fig.~\ref{fig:coh-superdense}).

The global state shared between Alice and Bob at the start of the protocol is
a four-mode state. Alice possesses two modes $A_{1}$ and $A_{2}$.\ She wants
to simulate a MQ conat channel on $A_{1}$ and a PQ conat channel on $A_{2}$.
Alice and Bob possess an $\epsilon$-position-correlated state shared between
modes three $\left(  \hat{x}_{A},\hat{p}_{A}\right)  $ and four $\left(
\hat{x}_{B},\hat{p}_{B}\right)  $ with mean-zero relative position and total
momentum. Alice performs a controlled-position displacement on her modes two
and three followed by a controlled-momentum displacement on modes one and
three. The controlled displacements replace the conditional displacements in
Braunstein and Kimble's protocol. Alice sends mode three to Bob via the qunat
channel. Bob performs an inverse controlled-position displacement on modes
three and four followed by reflecting mode four. The last two operations
replace the beamsplitter in the original dense coding protocol. The last two
operations are also the inverse operations of the preparation stage for the
coherent teleportation protocol (compare Fig.~\ref{fig:coh-tele-harrow} and
Fig.~\ref{fig:coh-superdense}). The four modes then have the
Heisenberg-picture observables:%
\begin{align}
\hat{x}_{1}  &  =\hat{x}_{A_{1}}-\left(  \hat{x}_{A_{2}}+\hat{x}_{A}\right)
,\ \ \hat{p}_{1}=\hat{p}_{A_{1}}\\
\hat{x}_{2}  &  =\hat{x}_{A_{2}},\ \ \hat{p}_{2}=\hat{p}_{A_{2}}-\hat{p}%
_{A}\nonumber\\
\hat{x}_{3}  &  =\hat{x}_{A_{2}}+\hat{x}_{A},\ \ \hat{p}_{3}=\hat{p}_{A_{1}%
}+\left(  \hat{p}_{A}+\hat{p}_{B}\right) \nonumber\\
\hat{x}_{4}  &  =\hat{x}_{A_{2}}+\left(  \hat{x}_{A}-\hat{x}_{B}\right)
,\ \ \hat{p}_{4}=-\hat{p}_{B}\nonumber
\end{align}
Modes one and three satisfy the conditions for an $\epsilon$-approximate MQ
conat channel. Modes two and four satisfy the conditions for an $\epsilon
$-approximate PQ conat channel. Coherent superdense coding thus gives an
operational interpretation to the PQ and MQ conat channels.%

Coherent teleportation and superdense coding for DVs are dual under resource
reversal \cite{prl2004harrow,devetak:140503}, meaning that the resources
generated by one protocol are consumed by the other and vice versa. Duality
under resource reversal is only possible \textit{to some degree} with CVs
because of finite squeezing. We first give two ways of composing the
protocols. We then illustrate how the duality does not hold for some finite
number of compositions.

Implementing coherent teleportation with coherent superdense coding is one way
of composition. Coherent superdense coding (Fig.~\ref{fig:coh-superdense}%
)\ plays the role of the coherent channels in coherent teleportation
(Fig.~\ref{fig:coh-tele-harrow}). Alice wishes to teleport a mode $A_{1}$ to
Bob. Suppose Alice and Bob share an $\epsilon_{1}$-position-correlated state
with mode operators $\left(  \hat{x}_{A},\hat{p}_{A},\hat{x}_{B},\hat{p}%
_{B}\right)  $ and an $\epsilon_{2}$-position-correlated state with mode
operators $\left(  \hat{x}_{\bar{A}},\hat{p}_{\bar{A}},\hat{x}_{\bar{B}}%
,\hat{p}_{\bar{B}}\right)  $. Both correlated sets have mean-zero relative
position and total momentum. They use the first correlated state for coherent
teleportation. They use the second correlated state for coherent superdense
coding---which in turn implements both an $\epsilon_{2}$-approximate MQ conat
channel and an $\epsilon_{2}$-approximate PQ conat channel. The results are
the same as coherent teleportation (Fig.~\ref{fig:coh-tele-harrow}) with some
modifications. Alice and Bob possess an $\left(  \epsilon_{1}+\epsilon
_{2}\right)  $-momentum-correlated state shared between modes one and three
and an $\left(  \epsilon_{1}+\epsilon_{2}\right)  $-position-correlated state
shared between modes two and four. Mode five is the teleported mode with
Alice's original state $A_{1}$. A lower bound on the fidelity $F$ using this
protocol is $2/\left(  2+\epsilon_{1}+\epsilon_{2}\right)  $. The fidelity
exceeds the limit of $1/2$ if both $\epsilon_{1}<1$ and $\epsilon_{2}<1$.

Implementing coherent superdense coding with coherent teleportation is another
way of composition. The alternate coherent teleportation in
Fig.~\ref{fig:coh-tele-slick} replaces the qunat channel in
Fig.~\ref{fig:coh-superdense}. The protocol is similar to coherent superdense
coding except for an additonal set of two correlated modes. It begins with
Alice possessing two modes represented by quadrature operators $\left(
\hat{x}_{A_{1}},\hat{p}_{A_{1}},\hat{x}_{A_{2}},\hat{p}_{A_{2}}\right)  $. She
wishes to implement a PQ and MQ conat channel on these two modes. Suppose that
Alice and Bob possess an $\epsilon_{1}$-position-correlated state represented
by the quadrature operators $\left(  \hat{x}_{A},\hat{p}_{A},\hat{x}_{B}%
,\hat{p}_{B}\right)  $ with mean-zero relative position and total momentum.
Alice performs controlled displacements on her three modes
(Fig.~\ref{fig:coh-superdense}). She uses alternate coherent teleportation to
implement the qunat channel. She uses an $\epsilon_{2}$-approximate PQ conat
channel and an $\epsilon_{3}$-approximate MQ conat channel for alternate
coherent teleportation. Bob performs the last two operations in
Fig.~\ref{fig:coh-superdense}. The six modes then have the Heisenberg-picture
observables:%
\begin{align}
\hat{x}_{1}  &  =\hat{x}_{A_{1}}-\left(  \hat{x}_{A_{2}}+\hat{x}_{A}\right)
,\ \ \ \ \hat{p}_{1}=\hat{p}_{A_{1}}\\
\hat{x}_{2}  &  =\hat{x}_{A_{2}},\ \ \ \ \hat{p}_{2}=\hat{p}_{A_{2}}-\hat
{p}_{A}\nonumber\\
\hat{x}_{3}  &  =\hat{x}_{A_{2}}+\hat{x}_{A}+\hat{x}_{\Delta_{P}}%
,\ \ \ \ \hat{p}_{3}=\hat{p}_{A_{1}}+\hat{p}_{A}+\hat{p}_{\Delta_{X}%
}\nonumber\\
\hat{x}_{4}  &  =\hat{x}_{A_{2}}+\hat{x}_{A}+\hat{x}_{\Delta_{X}}\nonumber\\
\hat{p}_{4}  &  =\hat{p}_{A_{1}}+\left(  \hat{p}_{A}+\hat{p}_{B}\right)
+\left(  \hat{p}_{\Delta_{X}}+\hat{p}_{B^{\prime}}\right)  +\hat{p}%
_{\Delta_{P}}\nonumber\\
\hat{x}_{5}  &  =\hat{x}_{B^{\prime\prime}}-\left(  \hat{x}_{A_{2}}+\hat
{x}_{A}+\hat{x}_{\Delta_{X}}\right) \nonumber\\
\hat{p}_{5}  &  =\hat{p}_{A_{1}}+\hat{p}_{A}+\hat{p}_{\Delta_{X}}+\hat
{p}_{\Delta_{P}}\nonumber\\
\hat{x}_{6}  &  =\hat{x}_{A_{2}}+\left(  \hat{x}_{A}-\hat{x}_{B}\right)
+\hat{x}_{\Delta_{X}},\ \ \ \ \hat{p}_{6}=-\hat{p}_{B}\nonumber
\end{align}
where Alice possesses modes 1-3 and Bob possesses modes 4-6. Alice and Bob
share an $\left(  \epsilon_{2}+\epsilon_{3}\right)  $-momentum-correlated
state between modes three and five. Alice implements an $\left(  \epsilon
_{1}+\epsilon_{2}+\epsilon_{3}\right)  $-approximate MQ conat channel between
modes one and four if $\epsilon_{1}+\epsilon_{2}+\epsilon_{3}<1$. She
implements an $\left(  \epsilon_{1}+\epsilon_{2}\right)  $-approximate PQ
conat channel between modes two and six if $\epsilon_{1}+\epsilon_{2}<1$.

Coherent teleportation and coherent superdense coding\ are dual under resource
reversal in the sense given by the above first-order compositions if the
$\epsilon$ quantities are small enough. But examine the above protocols to
observe a loss of duality. The $\epsilon$ quantities accumulate additively
when composing multiple protocols using imperfect conat channels. Repeated use
of nonideal conat channels eventually degrades the available level of
squeezing until the sufficient conditions for entanglement and for
teleportation fidelity exceeding the classical limit no longer hold. The
protocols are not dual under resource reversal after some maximum number of
compositions due to finite-squeezing losses.

We provided conat-channel definitions and demonstrated several coherent
protocols. We concluded with an analysis of the duality under resource
reversal of coherent teleportation and coherent superdense coding. The conat
channel should lead to other CV\ coherent protocols. We thank Igor Devetak for
useful discussions and Geza Giedke for a useful comment. TAB and HK
acknowledge support by NSF Grant CCF-0448658.

\bibliographystyle{apsrev}
\bibliography{conat}

\end{document}